\documentclass[12pt,a4paper]{article}
\usepackage{times}
\usepackage{graphicx,epsf}
\title{Magnetic Phase Transitions in the double spin-chains compound $\rm LiCu_2O_2$}

\author{B. Roessli\\
\small{Laboratory for Neutron Scattering, Paul Scherrer Institute and ETH Zurich,
CH-5232 Villigen PSI, Switzerland }\\
U. Staub\\
\small{Swiss Light Source, Paul Scherrer Institute,
CH-5232 Villigen PSI, Switzerland} \\
A. Amato, D. Herlach\\
\small{Laboratory for Muon-Spin Spectroscopy, Paul Scherrer Institute,
CH-5232 Villigen PSI, Switzerland}\\
P. Pattison\\
\small{Institut de Crystallographie, Universit\'e de Lausanne, CH-1015 Lausanne, Switzerland }\\
K. Sablina, G.A. Petrakovskii\\
\small{Institute of Physics SB RAS, 660036 Krasnoyarsk, Russia}}

\date{\today}
\begin{document}
\maketitle
\begin{abstract}
\indent    
We report high-resolution x-ray diffraction, muon-spin-rotation spectroscopic and
 specific heat measurements in
the double spin-chains compound $\rm LiCu_2O_2$. The x-ray diffraction
results show that the crystal structure of $\rm
LiCu_2O_2$ ~is orthorhombic down to T=10K. Anisotropic line-broadening of the
 diffraction peaks is observed, indicating
disorder along the spin chains.  Muon spin relaxation and specific heat
 measurements show that
$\rm LiCu_2O_2$ ~undergoes a phase transition to a magnetic ordered state at $\rm T_1\sim24K$.
The specific heat data exhibits a second $\rm \lambda$-like peak at 
$\rm T_2\sim22.5 K$, which increases with increasing magnetic field  
similarly way to that found in spin-ladder compounds.
\end{abstract}

\small{\textbf{
Corresponding author: \\}}
Bertrand Roessli\\
 Laboratory for Neutron Scattering, 
Paul-Scherrer Institut and ETH Zurich\\
 CH-5232 Villigen, PSI.\\
e-mail: Bertrand.Roessli@psi.ch

P.A.C.S.:76.75+i, 65.40, 75.10.Jm
Keywords: muon spin rotation, spin-ladder, magnetic ordering

\newpage

\section{Introduction}
One-dimensional S=$\rm 1\over2$ antiferromagnets have physical
 properties which can only be accounted for
by quantum effects. The ground-state and the spectrum of excitations 
of a Heisenberg spin-chain with
nearest-neighbors interaction are known exactly and the theoretical results 
are in good agreement with
experiments \cite{perring}. Compounds with coupled S=1/2 chains  
are the subject of intense investigations as
they represent intermediate structures between one- and two-dimensional 
compounds. In this class of materials,
antiferromagnetic long-range order has been observed for compounds with 
zig-zag chains like $\rm SrCuO_2$
\cite{matsuda} or with weak inter-chains exchange interactions like $\rm 
Sr_2CuO_3$ and $\rm Ca_2CuO_3$
\cite{kojima}. A common property of these materials is that both the size
 of the magnetic moments at
saturation and the N\'eel temperature are strongly reduced due to frustration between
exchange integrals and quantum fluctuations\cite{white}. In addition, understanding the magnetic
properties of coupled-chains
compounds is of interest for copper-oxides high-$T_c$ superconductors, 
which can be considered as two-dimensional
spin-$\rm 1\over 2$ antiferromagnets with carrier doping.

$\rm LiCu_2O_2$ is a mixed-valent compound with copper ions in the
 $\rm Cu^{2+}$ and $\rm Cu^{1+}$ oxidation
states\cite{zatsepin}. At first the chemical structure of 
$\rm LiCu_2O_2$ was described within the
tetragonal space group $P4_2/nmc$ \cite{hibble}. Later x-ray and neutron measurements \cite{berger} 
suggested that
$\rm LiCu_2O_2$ crystallizes in the orthorhombic space-group $Pnma$ 
with lattice constants a=5.72 \AA, b=2.86 \AA
~and c=12.4 \AA ~at room temperature. The chemical structure of
 $\rm LiCu_2O_2$ may be viewed as chains of
$\rm Cu^{2+}$ ions propagating along the b-axis. There are two such parallel Cu-chains which run 
along the a-axis and which are bridged along the c-axis
by a 90$^\circ$ oxygen bond, as shown in Fig.\ref{Fig1}. The double-chains
are well isolated from each other by both Li-ions and sheets of 
 non-magnetic $\rm Cu^{1+}$ ions. From these considerations, it
appears that $\rm LiCu_2O_2$ is a good candidate for either a 
spin-ladder or a zig-zag chain system, depending on
the ratio of the nearest- to second-nearest neighbor exchange interactions. 
In this paper, we report high-resolution x-ray powder diffraction,
muon-spin-rotation spectroscopy ($\rm \mu SR$) and specific heat 
results in the spin-$\rm 1\over 2$ chain-like compound $\rm LiCu_2O_2$. 
The results suggest that antiferromagnetic ordering is induced by chemical 
disorder along the chains.
\\
\section{Experimental results and Discussion}
Single-crystals were prepared by the spontaneous crystallization
 method starting from $\rm Li_2CO_3$ and $\rm
CuO$. A detailed description of the preparation method is given elsewhere
\cite{vorotynov}. For the experiments reported here, single-crystals of typical size $\rm 3 \times
3 \times 1 mm^3$ were used. Special care is to be taken because
$\rm LiCu_2O_2$ ~oxidizes in open air. Therefore, the samples were kept under
 dry helium atmosphere. X-ray
diffraction with Cu $\rm K_\alpha$ radiation showed that the single crystals
 contain traces of $\rm Li_2CuO_2$
($\sim$3\%). The structural and magnetic properties of $\rm Li_2CuO_2$ are well established
\cite{li2} and can easily be separated from those of $\rm LiCu_2O_2$.
\\
The high-resolution x-ray diffraction experiments were performed at the Swiss-Norwegian 
Beam line at the
European Synchrotron Radiation Facility (ESRF) in Grenoble, France. A 
diffraction
 Debye-pattern was
collected in standard Scherrer geometry with a wave length of $\rm \lambda$=0.49876 \AA ~at room
temperature. The 2$\rm \theta$
resolution was improved to 0.01$^\circ$ by means of four Si (111) analyzer crystals.  For
the low temperature experiments a $\rm ^4He$-flow cryostat was installed to cool the sample down to
10~K. The experiments at low temperatures were performed with $\rm \lambda$=0.79764 \AA.
Fine powder of $\rm LiCu_2O_2$ was sealed in a 0.3 mm diameter quartz capillary.
The $\rm \mu SR$ experiments were performed on the GPS spectrometer at the Paul-Scherrer Institut,
Switzerland. The sample 
consisted of approximately 10 crystals which were glued on a silver plate with the
crystallographic c-axis oriented along the muon path. A
$^4$He flow-cryostat  was used to obtain temperatures between 10K
$\le T \le$ 30K. The calorimetric measurements were performed with a 
commercial PPMS (Quantum Design) device
in the temperature range  1.8K$\le T \le$100K. 
Fig.\ref{Fig2} shows the splitting of the 400 and 200 reflections
determined at T= 10K. This splittong is a direct evidence 
 that the chemical structure of $\rm LiCu_2O_2$ is
orthorhombic. Diffraction reflections
with Miller indices $\rm h,k\ne 0$ are found to be broader than those 
with h=0, k=0. This anisotropic line-broadening indicates atomic disorder in 
the crystallographic (a,b)-plane. 
The in-plane correlation length, as calculated from the half-width at half 
maximum of the Bragg peaks, amounts to
$\sim$540\AA. The orthorhombic strain
$\rm (a^*-b^*)/(a^*+b^*)$ distinctly decreases from room temperature
 to T=10K by a factor of $\sim2$,
which is not expected (see Fig. \ref{Fig3}). Usually, 
materials exhibit the tendency to approach higher
symmetries for increasing temperatures as the increase of 
the lattice vibrations as a function of temperature leads to 
the relaxation of the lattice and to the reduction of the strain. 
From an extended diffraction pattern taken at room
temperature, we conclude that the chemical structure of
$\rm LiCu_2O_2$ is well described  with the space group $Pnma$ and lattice
constants a=5.7301(2)~\AA, b=2.8594(1)~\AA ~and c=12.4192(3)~\AA. The understanding of the temperature dependence of the strain, however, requires a more detailed structural studies at elevated temperatures. 
\\
In $\rm \mu SR$ experiments the asymmetric emission of positrons arising from the weak decay of 
implanted spin-polarized muons is monitored. The time-dependent
positron rate $N(t)$  is recorded as a function of time
and is given by the function 
\begin{equation}
N(t)=N(0)\exp(-t/\tau)[1+AG_z(t)]~,
\label{rate}
\end{equation}
where $A$ is the initial muon asymmetry parameter, $G_z(t)$ the asymmetry function 
and $\rm \tau$ is the muon life-time. 
Zero-field $\mu$SR signals in $\rm
LiCu_2O_2$ are shown in
Fig.\ref{Fig4}.
At T=28 K, the asymmetry function does not reveal frequencies, indicating that the
magnetic moment of the implemented muon does not undergo Larmor precession.
In this temperature regime, the muon
spin depolarisaton originates from the magnetic fields  caused by the Cu nuclear dipole
moments.
Assuming that these internal fields have a Gaussian distribution 
and that they are randomly oriented, the
asymmetry function
$G_z$(t) is given by the familiar Kubo-Toyabe expression
\cite{Kubo} 
\begin{equation}
G_{KT}(t)=\frac{1}{3}+\frac{2}{3}(1-\Delta^2t^2)
\exp(-\frac{1}{2}\Delta^2t^2)
\label{rns}
\end{equation}
where $\Delta^2/\gamma^2_{\mu}$ represents the second moment of the field distribution 
 and $\gamma_{\mu}$ = 2$\pi\cdot$13.553879 kHz/G is the giromagnetic
 ratio of the muon.
At T=28~K, a fit to the $\rm \mu SR$ data yields a dipolar width 
$\Delta$=0.324 (MHz). Upon lowering the
temperature below $\rm T_1\sim 24K$, a precession of the muon spins is observed. 
This suggests that
the $\rm Cu^{2+}$ spins develop a static magnetic order below that temperature.
A characteristic feature
of the muon signal determined in $\rm LiCu_2O_2$ at low temperatures is that  
the frequencies show a damping as a function of increasing decay time, 
which indicates a broad distribution of
 magnetic fields at the muon stopping sites.
\noindent 
In the temperature
range
$10\rm K\leq T\leq 24\rm K$, the data are best described by assuming for
$G_z(t)$ the form
\begin{eqnarray}
G_{z}(t)&=&A_1\exp[-(\lambda t)] \nonumber \\
&+& A_{f1}e^{-\gamma_1t}cos(2\pi\omega_1t+\phi)\nonumber \\
&+& A_{f2}e^{-\gamma_2t}cos(2\pi\omega_2t+\phi).
\label{osc}
\end{eqnarray}
$\phi$ is given by the position of the positron detectors relative to the muon polarization.
The first term of Eq.
\ref{osc}, which arises from the non-zero projection of the muon-spin
 polarization along the direction of the internal
fields, indicates the presence of fast longitudinal  fluctuations 
in $\rm LiCu_2O_2$. This suggests that even 
in the ordered  magnetic phase
the $\rm Cu^{2+}$ magnetic moments are not fully static.
A least-square fit to the muon data in the temperature range $\rm 10K \le T \le 23K$ 
with Eq. \ref{osc} yields essentially temperature independent parameters apart from the Larmor
frequencies. They  show a dependence as a function of temperature reminiscent
 of static order parameters measured in ordered ferro- or antiferromagnets (Fig. \ref{Fig5}).
The fitted values for the relaxation rates are
 $\rm \lambda=0.14 ~MHz, \gamma_1\sim 5.5 ~MHz$ and $\rm \gamma_2\sim 7 ~MHz$, respectively.
In the temperature range $\rm 22K \le T \le 24K$, the relaxation rate $\rm \lambda$ increases
which indicates that the magnetic moments in $\rm LiCu_2O_2$ fluctuate 
faster when approaching the ordering
temperature $\rm T_1=24K$. 
The fact that the damping of the muon spin precession is temperature 
independent below $\rm T \le 19K$ indicates that 
some static inhomogeneity in the $\rm Cu^{2+}$ magnetic moments along 
the spin-chains is present in $\rm LiCu_2O_2$. A similar situation is 
encountered in Zn- and Si-doped
$\rm CuGeO_3$, where it has been shown that 
impurities along the spin-chains result in spatial 
variation of the size of the magnetic moments around 
the doping center \cite{kojima2}. 
Accordingly, staggered moments will be induced along the spin-chains 
which eventually leads to 
static N\'eel order\cite{fukuyama}. This point of view has been adopted in a previous study of  
the magnetic properties of $\rm LiCu_2O_2$ by magnetic susceptibility and 
resonance measurements\cite{vorotynov}. The magnetic susceptibility of $\rm LiCu_2O_2$ shows a broad
maximum around $\rm T\sim 50K$. As the temperature dependence of the
 magnetic susceptibility is well
reproduced with a Heisenberg model for interacting chains\cite{vorotynov,fritschij}, it 
has been concluded that
$\rm LiCu_2O_2$ is a low-dimensional system of spin-ladder type.
 A $\rm S={1\over 2}$ spin-ladder structure has
a singlet ground-state\cite{rice} and as such does not develop 
long-range order. However, antiferromagnetic
resonance lines were observed in $\rm LiCu_2O_2$ by the authors of 
Ref.\cite{vorotynov} below T=22.5K.
It was therefore argued that $\rm LiCu_2O_2$ is an antiferromagnet 
below $\rm T_N\sim 22.5K$ as a consequence
of partial redistribution of copper and lithium ions along the chains. 
On the other hand,  antiferromagnetic ordering in 
$\rm LiCu_2O_2$ might also be due to small interchain interactions 
$\rm J_{perp.}$.
In that context, mean-field theory \cite{schultz} applied to the special of 
coupled-chains predicts a N\'eel temperature 
$\rm T_N$ proportional to $\rm J_{perp.}$. In the absence 
of precise exchange interactions between the $\rm Cu^{2+}$ 
for $\rm LiCu_2O_2$, it is however 
difficult to draw definite conclusions at this stage 
about the origin of antiferromagnetic ordering in $\rm LiCu_2O_2$. 
Inelastic neutron measurements are therefore 
desirable. 
The $\rm \mu SR$ results clearly
indicate that $\rm LiCu_2O_2$ undergoes a phase transition to an 
ordered magnetic state below $\rm T_1$=24K.
A close look at the temperature dependence of the Larmor 
frequencies shown in Fig.\ref{Fig5} 
also reveals another anomaly around $\rm T_2$=22.5K.
The specific heat results are shown in Fig.\ref{Fig6}. A double peak structure
is observed in the temperature dependence of the specific heat with 
maxima at $\rm T_2=$22.5K and $\rm T_1$=24K. 
The specific heat of
$\rm LiCu_2O_2$ exhibits sharp $\rm \lambda$-like peaks which have different 
dependencies as a function of applied
magnetic field. The peak at $\rm T_1$=24K shifts to lower temperatures 
when a field is applied and is therefore
associated to the magnetic phase transition. On the other hand the 
$\rm T_2$=22.5K peak does not exhibit any
temperature shift but significantly increases as a function of
 magnetic field. \\
In conclusion, we have presented high-resolution x-ray powder diffraction,
 $\rm \mu SR$ and specific heat measurements in
$\rm LiCu_2O_2$. The data are consistent with the view that $\rm LiCu_2O_2$ can be considered as
a double-chains $\rm S={1\over 2}$ system for which N\'eel ordering below $\rm T_N=24K$ 
is induced by chemical
disorder along the spin-chains. The field dependence of 
the specific heat data shows features similar to the ones
observed in spin-ladder compounds.
\section{Acknowlegments}
The help of P. Allenspach with the specific heat measurements is gratefully acknowledged. 
It is a pleasure to thank the Swiss-Norwegian beam-line staff at ESRF for 
support..

\newpage
\begin{figure}
\centering
\vskip 4pt
\caption{Chemical structure of $\rm LiCu_2O_2$ showing the double $\rm Cu^{2+}$ chains.}
\label{Fig1}
\end{figure}
\begin{figure}
\centering
\vskip 4pt
\caption{(left) Selected reflections of the x-ray pattern taken at T=20K 
in $\rm LiCu_2O_2$ which shows the anomalous
broadening of the (210) Bragg reflection. 
The 006 peak was shifted by $\rm \sim$0.5$^\circ$ to obtain a superposition 
fo these reflections. 
(right) 400 and 020 reflections
reflecting the orthorhombic distortion. The lines are fits with a Voigt function.}
\label{Fig2}
\end{figure}
\begin{figure}
\centering
\vskip 4pt
\caption{Orthorhombic strain determined in $\rm LiCu_2O_2$. 
See text for details.}
\label{Fig3}
\end{figure}
\begin{figure}
\centering
\vskip 4pt
\caption{Experimental zero-field $\mu$SR signal measured in $\rm LiCu_2O_2$ on GPS. 
The line is the result
of a fit with the model function explained in the text.}
\label{Fig4}
\end{figure}
\begin{figure}
\centering
\vskip 4pt
\caption{Larmor frequencies observed in $\rm LiCu_2O_2$ on GPS. The lines are guides to the eyes.}
\label{Fig5}
\end{figure}
\begin{figure}
\centering
\vskip 4pt
\caption{Specific heat data measured in $\rm LiCu_2O_2$. The 
inset depicts the dependence of the $\rm
\lambda$-like peak for increasing and decreasing magnetic 
fields at $\rm T_2$=22.5K.}
\label{Fig6}
\end{figure}

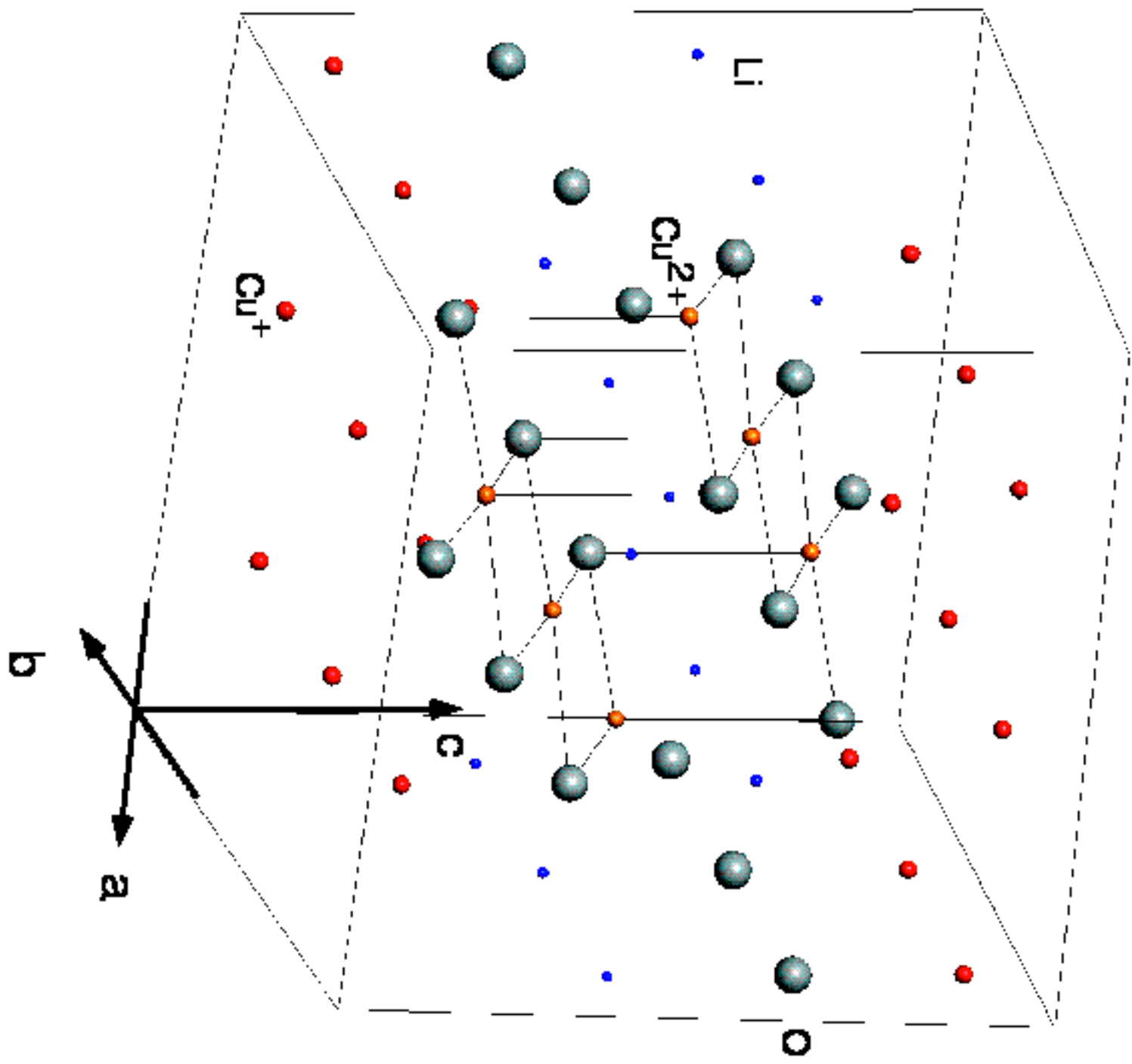
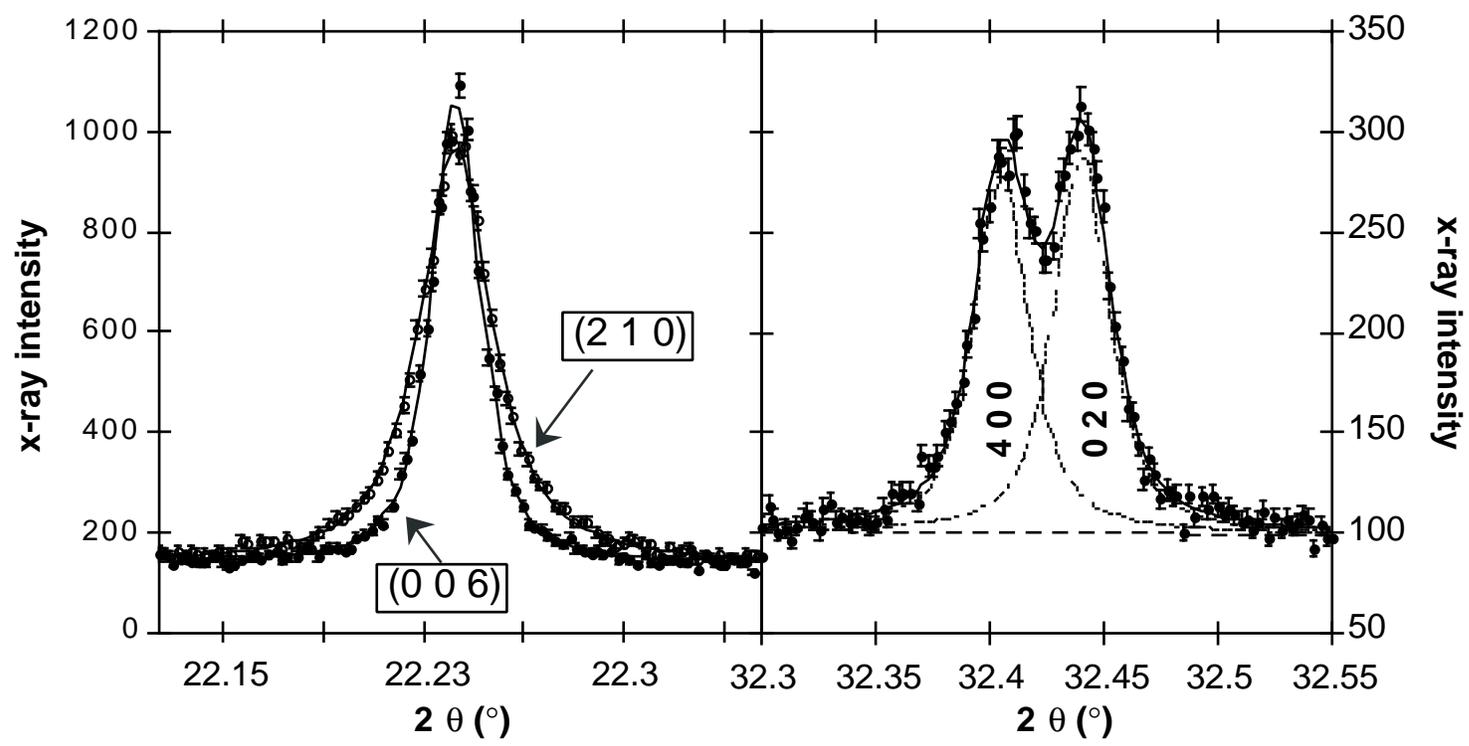
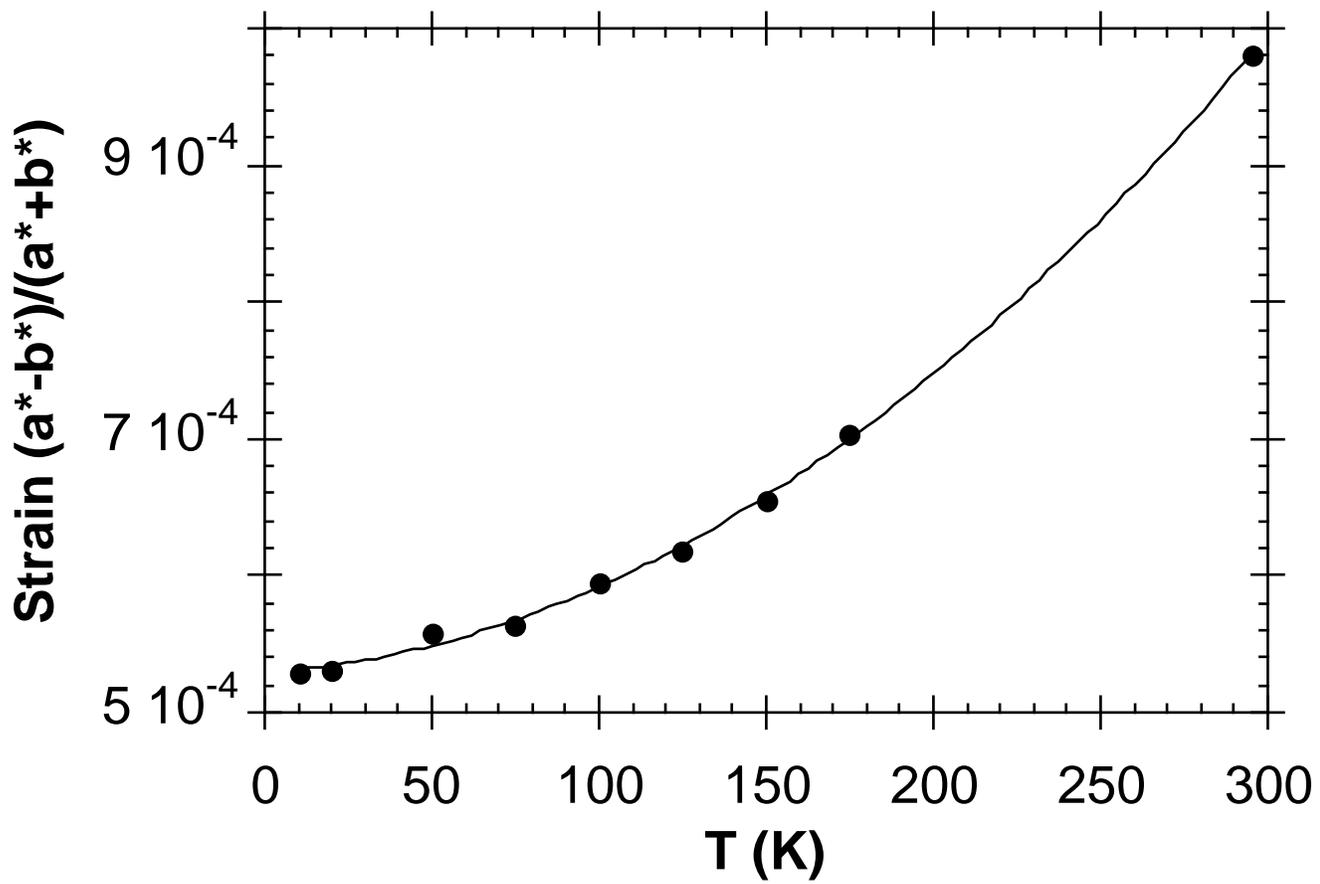
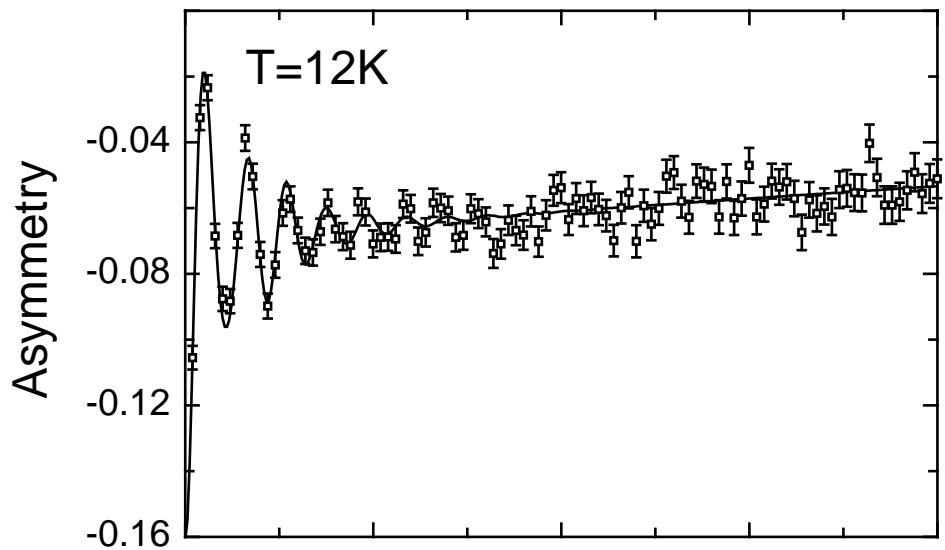
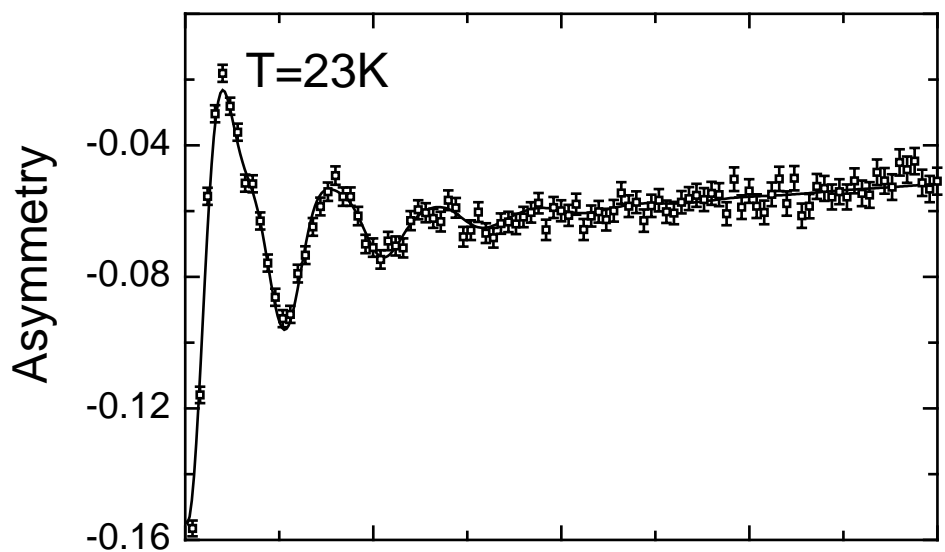
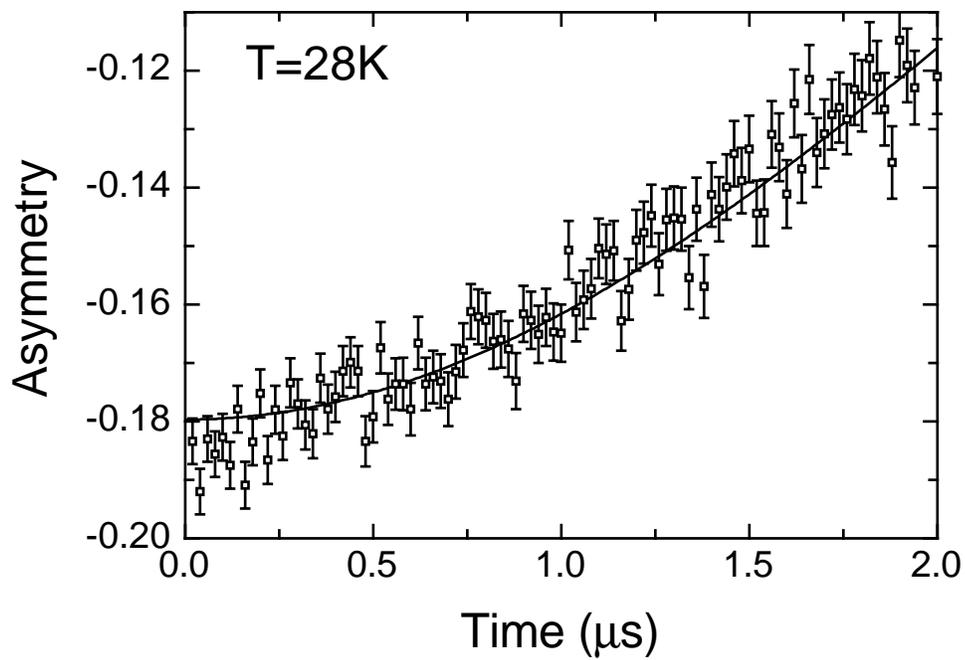
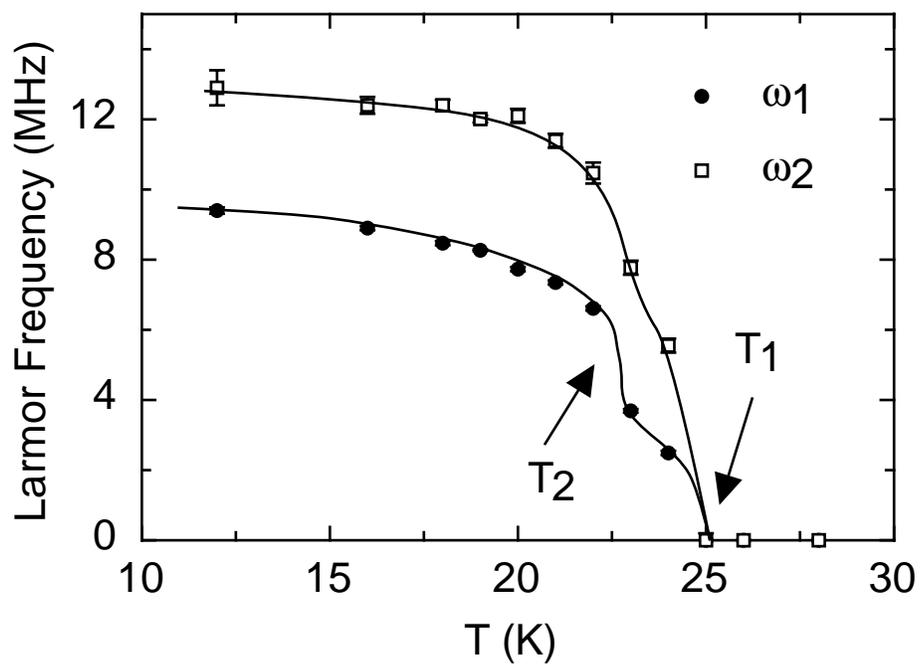
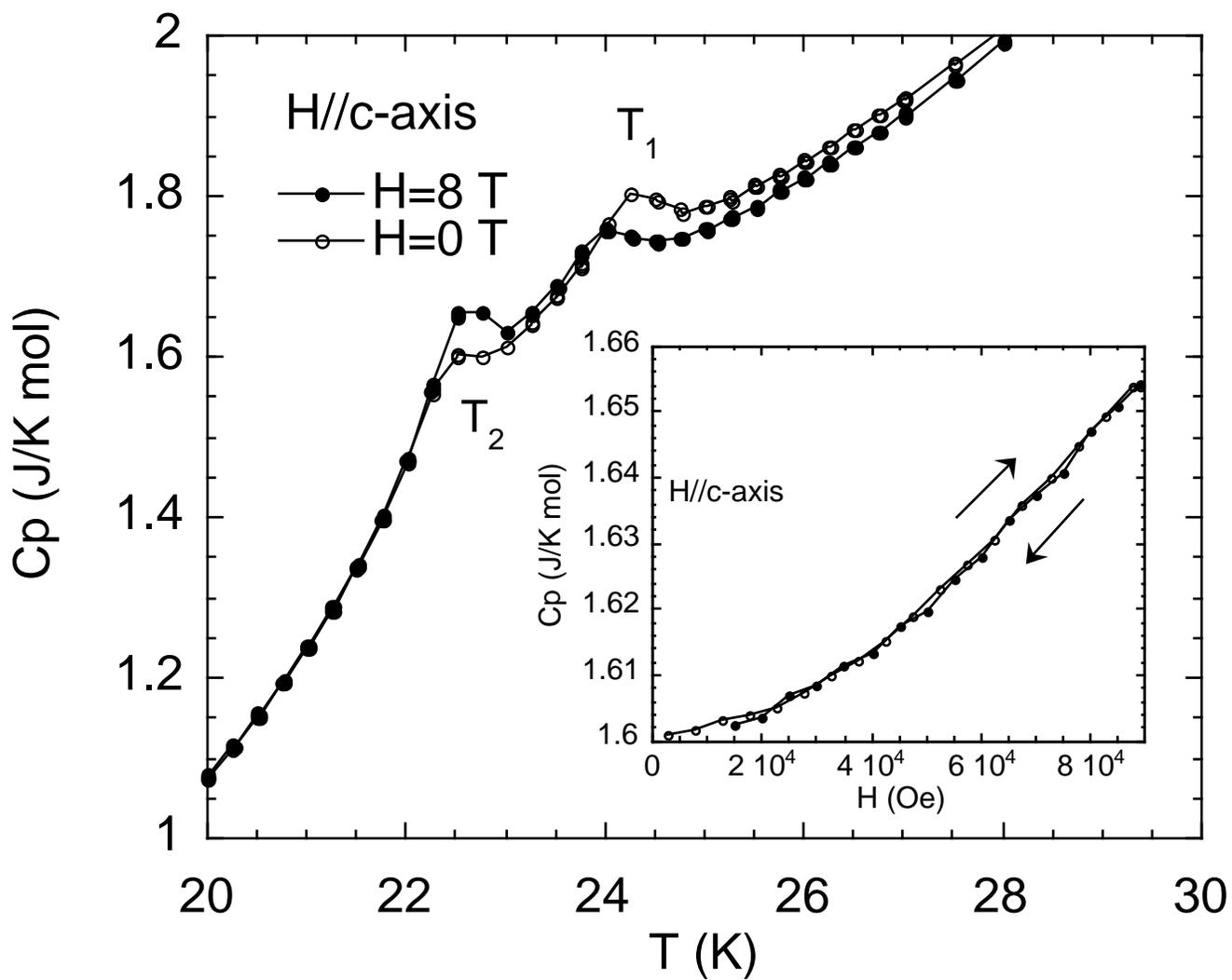
\end{document}